# Enrichment of gut microbiome strains for cultivation-free genome sequencing using droplet microfluidics

Anna Pryszlak[1,*], Tobias Wenzel[1,2,*], Kiley West Seitz[1], Falk Hildebrand[1,3,3a], Ece Kartal[1], Marco Raffaele Cosenza[1], Vladimir Benes[1], Peer Bork[1,4a,4b,+] and Christoph Merten[1,5,+]

[1] European Molecular Biology Laboratory, Heidelberg, DE
[2] Institute for Biological and Medical Engineering, Schools of Engineering, Medicine and Biological Sciences, Pontificia Universidad Católica de Chile, Santiago, CL
[3a] Gut Microbes and Health, Quadram Institute Bioscience, Norwich, UK
[3b] Digital Biology, Earlham Institute, Norwich, UK
[4a] Max Delbrück Centre for Molecular Medicine, Berlin, DE
[4b] University of Würzburg, Würzburg, DE
[5] School of Engineering, Institute of Bioengineering, École Polytechnique Fédérale de Lausanne, Lausanne, CH
* Both authors contributed equally to this work
+ Correspondence: christoph.merten@epfl.ch and peer.bork@embl.org.

## Abstract

We report a droplet microfluidic method to target and sort individual cells directly from complex microbiome samples, and to prepare these cells for bulk whole genome sequencing without cultivation. We characterize this approach by recovering *bacteria* spiked into human stool samples at a ratio as low as 1:250 and by successfully enriching endogenous *Bacteroides vulgatus* to the level required for de-novo assembly of high-quality genomes. While microbiome strains are increasingly demanded for biomedical applications, the vast majority of species and strains are uncultivated and without reference genomes. We address this shortcoming by encapsulating complex microbiome samples directly into microfluidic droplets and amplify a target-specific genomic fragment using a custom molecular TaqMan probe. We separate those positive droplets by droplet sorting, selectively enriching single target strain cells. Finally, we present a protocol to purify the genomic DNA while specifically removing amplicons and cell debris for high-quality genome sequencing.

## Introduction

The microbiome is a complex and increasingly important research target. In particular the gut microbiome is linked to over 100 disease states or syndromes[1–4], and contains at least several thousand species[5]. A person's microbiome influences how its body metabolises drugs[6,7] and therefore has an impact on success or failure of any pharmacological intervention. Current microbiome studies are mainly driven by innovations in the analysis of metagenomic sequencing data. It has, for instance, recently become possible to reconstruct metagenomically assembled genomes (MAGs) of abundant taxa in metagenomic datasets[5,8,9]. However, beyond abundance limitations, MAGs have been found to be prone to chimeric assemblies[10]. Metagenomic approaches are also limited in resolving mobile elements, e.g. showing which species share the same genes due to horizontal gene transfer, an information of increasing relevance in medical research on antibiotic resistance and toxin encoding genes.

More complementary experimental high-throughput approaches are needed to perform "precision genomics". Beyond the benefits of increased resolution through targeted experiments[11], microfluidic droplet screens are able to provide the required throughput to study complex ecosystems such as microbiomes[12]. Experiments in droplets can also substantially lower the amount of material and reagents used, cost and contamination[13,14], and provide additional information through functional enrichment[15,16]. An increasing diversity of microfluidic tools for bacterial single cell analysis exists in the context of laboratory cultures. However, few natural isolates from complex microbiomes are available as lab cultures and most species have never been cultured at all[5,17].

In terms of cultivation-free techniques for the genomic analysis of microbiota cells, it has been achieved to generate single amplified genomes (SAGs) in microfluidic gel-microdroplets[18] or fused emulsions[19], with barcoding either inside droplets or after isolation in wells. This approach has yielded novel information on uncultivated and low-abundance strains at a single-cell level with the trade-off of low genome coverage and amplification biases. SAG generation is not targeted and would not allow to specifically look for microbes carrying particular genes of interest. Genetic targeting has been demonstrated by earlier studies performing polymerase chain reaction (PCR) in droplets to generate a sequence dependent florescent signal, used for PCR activated droplet sorting (PADS)[20,21]. This was shown for plasmids and phage sequences in *E.coli* laboratory cultures, and more recently for genetic libraries cloned into *E.coli*[22]. However, it was never applied to complex microbiomes of unknown, highly diverse biochemical composition, additionally hindered by sample contaminants such as fibres and particles found in stool samples.

To combine strengths of culture-free approaches and hypothesis-driven targeted approaches, we have established a genomics resource where a desired strain is enriched without cultivation and where genome fragments are sequenced in bulk to recover high quality genomes. We demonstrate a workflow that directly processes complex medically relevant microbiome samples, which are far less homogeneous and defined than laboratory cultures. By testing directly on complex samples, we aim to make our development a useful resource for the microbiome research community. While building on FADS methods, we show here how to apply them directly to microbiome samples, how to target bacteria without reference genome, how the method can be used to estimate the absolute abundance of a particular species, and how the abundant amplicon sequences used for droplet identification can be removed prior to genome sequencing. Our resource's workflow is summarised in Figure 1.

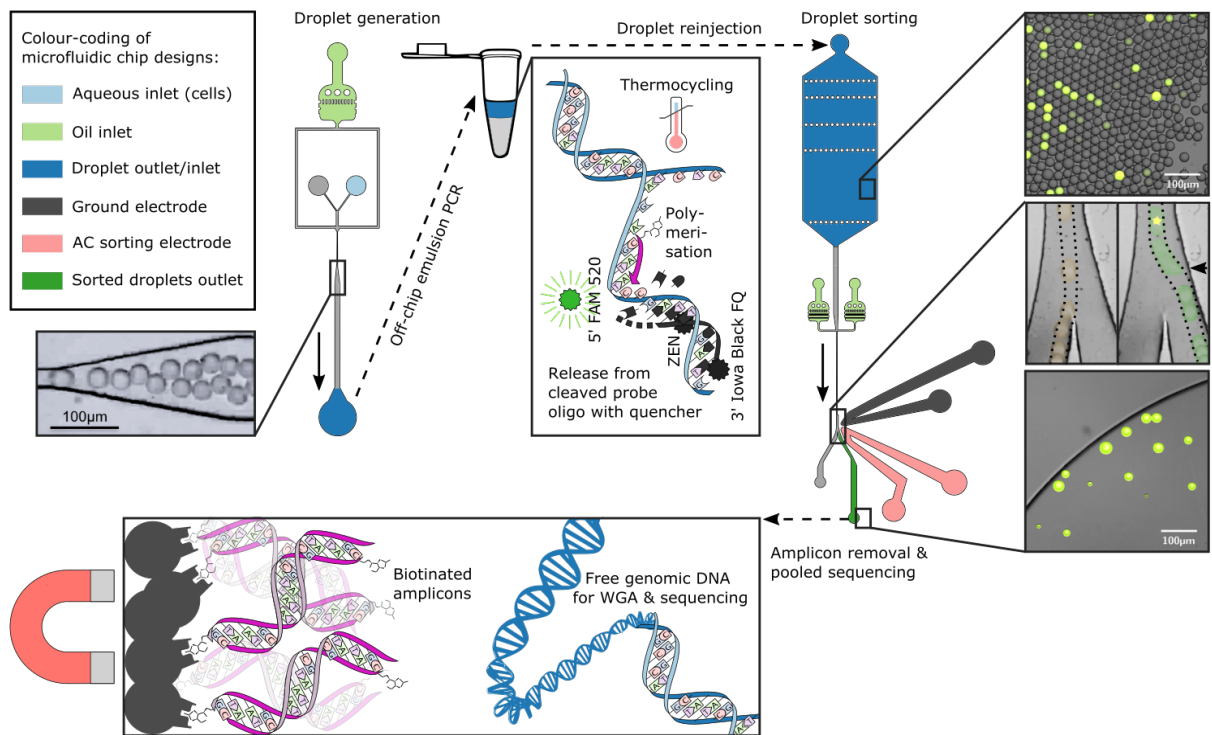

*Figure 1*: Microfluidic chip designs, microscope images of their use, and schematics of molecular mechanisms. ***Top-left***: Droplet generation chip and photo of 39µm-droplet generation with cells and reagents in oil. The arrow indicates the flow direction. ***Top-center***: Illustration of emulsion PCR in a thermocycler. If the target DNA sequence is present, it allows the binding of biotinylated primers and probes and thus the synthesis of corresponding amplicons. During strand extension the TaqMan probes are cleaved, releasing fluorescent molecules. ***Top-right***: Droplet sorting chip design and microscopy images. The microscopy pictures on the right show an emulsion before sorting, during sorting, and the positive droplet enrichment post-sorting. Here, fluorescent droplets indicate the presence of the gut bacterium *B. vulgatus* inside the droplet. Multiple microscopy images of the droplet sorting junction at different timepoints were overlaid and coloured to demonstrate droplet flow-traces. In the right image, a small black arrow indicates the location of the nearby sorting electrode, and a bright green spot indicates the upstream location of fluorescence detection. The scale bars correspond to 100µm. ***Bottom:*** Illustration of the removal of abundant amplicons with biotin-binding Streptavidin beads (after pooling the sorted positive droplets) to purify genomic DNA for sequencing library preparation. Library preparation further involves whole genome amplification (WGA) after binding adapter primers to the randomly broken genomic DNA fragments, followed by DNA fragment size selection.

## Materials and Methods

**Stool sample preparation and cell counting:** 0.5ml stool sample (measured using a volume scaled syringe cylinder) was dissolved in 4.5ml of filtered (0.22µm) 0.9% saline solution. The sample was then further diluted to obtain a final dilution of 1:50. Sterilised glass rattler beads (generic) were added and samples were homogenized on a Vortex mixer for 1min in order to suspend the sample and detach microbiota from dietary fibres. The homogenization step was repeated until faecal pellets were completely dissolved. Suspension was filtrated through a 40µm Cell Strainer (Falcon) to remove all remaining large organic particles and clumps, using gravity flow only. Aliquots were prepared from this suspension and frozen at -80°C.

Flow cytometric cell counting of stool and cell cultures was performed with a modified protocol of van der Waaij et al.[23], using bacteria staining with SYTO® BC and reference beads to assess the measured liquid volume, following the suppliers protocol (Invitrogen B7277 Bacteria Counting Kit for flow cytometry). Flow cytometry was performed on a LSR Fortessa bench top analyser (BD), using the 488nm laser line and the 530/30-A detector.

**Primer and probe design and testing:** Primers and probes were designed with the Primer3 software (http://bioinfo.ut.ee/primer3-0.4.0/) and the National Center for Biotechnology Information (NCBI, https://www.ncbi.nlm.nih.gov/tools/primer-blast/) primer designing tools based on the mOTUs V2[24] marker gene sequences of the species of interest. Probes were designed using the same tools and PrimerQuest (https://www.idtdna.com/pages/tools/primerquest). To work well in this protocol, the annealing temperature should ideally be 60°C for the primer and 70°C for the probe (the length of amplicon should ideally be 80-150bp long), which was further checked with various online in-silico PCR simulation software tools. Candidate sequences were also extensively specificity-checked by blasting against large sequence databases of Microbiota (NCBI and internal). Several primer candidates were ordered (Sigma-Aldrich) and efficiency and specificity tested in a qPCR assay in triplicates against a number of purified DNA pools (human DNA, mix of E.coli strains, mix of 31 gut microbiota strains, DNA from stool samples), all diluted and aliquoted. We added 0.5ng of DNA to each reaction well, which corresponded to optimal signals in our experiments. qPCR tests were performed with Syber green master mix including ROX reference dye. The most efficient primer pairs free of PCR side products and false-negatives where product-confirmed by gel-electrophoresis and Sanger sequencing of amplicons (Eurofins) and ordered again (from Sigma-Aldrich), this time with biotinylation for the actual experiments. TaqMan probes were purchased at IDT (sequence with 3' Iowa Black FQ, intermediate ZEN quencher and 5' 6-FAM fluorophore). Final primers and probes were tested in wells as well as droplets, using the PCR mix in Table 2. The best performing sequences for target organisms in this study are shown in Table 1.

*Table 1: Primers and probes sequences*

| Name | DNA sequence | Modification |
|---|---|---|
| ***Bacteroides vulgatus* and *dorei*** | | |
| Forward primer | CAAGCTGAGAAAAGCAGCCAAA | 5' Mod: BtnTg |
| Probe sequence | AGTGGCAGTAGCCGGAGGGGTATCAGCCA | /56-FAM/AG TGG CAG T/ZEN/A GCC GGA GGG GTA TCA GCC A/3IABkFQ/ |
| Reverse primer | GAATGAGTTACGAAGCCCGTTG | 5' Mod: BtnTg |
| ***Bacillus subtilis*** | | |
| Forward primer | TCGTGCTGAGACAGTTGCTT | 5' Mod: BtnTg |
| Probe sequence | TTGCGGGCGGCGGTATGGCAGGAGCT | /56-FAM/TTG CGG GCG /ZEN/GCG GTA TGG CAG GAG CT/3IABkFQ/ |
| Reverse primer | TCTTTCCCTTCAAGGCGGAC | 5' Mod: BtnTg |

**Preparation of microfluidic chips:** Two microfluidic devices, as shown in Figure 1, were drawn in AutoCAD (Autodesk) and printed black on UV-transparent polymer film (by SelbaTech) at a resolution of 25400 dpi. Master moulds of the microfluidic chips were then fabricated using soft-lithography on 100mm diameter silicon wafers (Silicon Materials) with the negative photoresist SU-8 2025 (MicroChem) and a mask aligner (Suess MicroTec MJB3). The droplet generation and sorting chips were produced with a channel height of 35um, while the sorting chip is a two-layer design with a taller droplet inlet chamber of 100um (using photoresist SU-8 2075, MicroChem). After standard baking and developing steps, a degassed PDMS and curing agent mix (Sylgard 184, Dow Corning Corp.) was poured onto the master moulds and cured overnight at 65°C. The silicone elastomer chips were then peeled off the master, cut and punched with biopsy punches (Harris Unicore) for tubing ports, and then bonded to microscopy glass slides (Thermo Fisher Scientific Inc.; droplet generation chips) or onto ITO glass (Delta Technologies LTD; ground electrode for sorting chips), after surface activation in an oxygen plasma oven (Femto, Diener electronic GmbH). The electrode channels of the sorting ships were filled with Indium-solder (0.5mm Solder Wire Indalloy19, Indium corporation of America) on a hotplate at ca. 85°C. Finally, before use, microfluidic channels were rinsed with Aquapel (Autoserv, Germany) and dried to increase hydrophobicity. Polytetrafluoroethylene (PTFE) tubing (Adtech Polymer Engineering Ltd & APT Advanced Polymer Tubing GmbH) was used to connect syringes containing liquids to inlet ports on the chips.

**Single-cell encapsulation into droplets and digital PCR:** Stool sample aliquots, TATAA Probe GrandMaster Mix (TA02-625), primer-mix, probe and dNTP-mix aliquots were defrosted and mixed according to Table 1. Prior to mixing the 800µl volume, the stool sample aliquot was washed to remove potentially free-floating DNA contaminants. To do so, the same volume of PBS buffer was added to the cell suspension, vortexed, and pelleted in a centrifuge for 12min at 3000g, after which the supernatant was discarded. The pellet was finally resuspended in PBS buffer corresponding to the original suspension volume by pipetting up and down 20 times. The PCR mix was optimized for best amplification efficiency in the droplets taking into account surfactant presence and the small volume of the reaction.

*Table 2: Encapsulation PCR mix with cells*

| Aqueous reagent | Volume | Concentration / Comment |
|---|---|---|
| Cell suspension aliquot | e.g. 50 µl | Volume adjusted according to cell suspension aliquot density to encapsulate on average one cell in every fourth or fifth droplet (ca. 30pl/droplet) |
| 10µM primer mix | 55 µl | 0.69µM final concentration of each primer (forward and reverse, aliquots mixed in advance at 10µM of each primer) |
| 100µM TaqMan probe | 20 µl | 2.5µM final concentration |
| Taq hotstart polymerase | 2 ul | To improve efficiency of late PCR cycles |
| 100µM mix of each dNTP | 6 µl | 0.75µM final concentration |
| Master mix | 440 µl | |
| Nuclease-free Water | Until 800 µl | |

The aqueous phase containing cells and PCR reagents and the oil phase (Bio-RAD droplet generation oil "for probes" or "for EvaGreen", depending on batch quality) were injected into the droplet generation chip using syringe pumps (Harvard apparatus). Droplets were produced within 60 min by flow focusing at flow rates of ca.1000µL/h for the aqueous phase and two-fold higher oil flow rates (with a droplet generation rate of ca. 8kHz). Droplets were collected and thermocycled off Chip in a 96 well plate (low profile skirted, Bio-RAD) according to the settings in Table 3, followed by storage at 37°C in the dark overnight (up to five days) for further downstream sorting procedures. Droplets without cells and with *B. subtilis* culture instead of stool were processed in parallel as quality control.

*Table 1: Droplet PCR thermocycle setting*

| Time | Temperature | Conditions |
|---|---|---|
| 95°C | 15 min | Hold for cell lysis[25], from here on sample biosafety level is reduced to L1. |
| 95°C | 3 sec | 45x |
| 60°C* | 16 sec | |
| 72°C | 2 min | Hold |
| 37°C | infinite | |
| * *Each program needs to be adjusted to probes/primers annealing temperature.* | | |

**Microfluidic fluorescence-activated droplet sorting:** In order to specifically enrich fluorescent droplets, sorting was performed with a set-up as previously described[26,27]. Droplets were reinjected into the sorting device at a frequency of ca. 100 – 250Hz (flow rates 15 – 30µl/h) using syringe pumps (Harvard apparatus) along with two spacer sheath oil inlets (ddPCR™ Droplet Reader Oil, Bio-RAD) at twenty times higher flow rates. To detect the fluorescent 6-FAM signal coming from cleaved TaqMan probes during the PCR reaction, droplets were excited using a blue diode laser (488nm, 20mW; Melles Griot) and the fluorescence intensity was measured at and upwards of the emission wavelength of 514nm using a PMT (Hamamatsu). A custom LabVIEW software was used to enable dynamic adjustments of PMT gain (0 – 1V), droplet fluorescence width and intensity thresholds for sorting, electrode voltage (1 – 1.5kV), AC pulse frequency (30 – 40 kHz), pulse duration and delay.

**Microscopy and evaluating positive droplet ratios:** Fluorescence microscopy was performed to analyse droplet fluorescence. Microscopy slides (Thermo Fisher Scientific) were wetted with Droplet Generation Oil (Bio-RAD) and droplets were imaged directly on slides or inside broad microfluidic imaging channels on slides using a 10-fold objective with a Nikon Ti-E widefield microscope. Images in the brightfield and green fluorescence channels were taken sequentially.
For the analysis, an ImageJ macro was developed (https://gitlab.com/marco.r.cosenza/simple_droplet_tools) to support the measurement of droplet fluorescence intensity.

**Amplicon removal and sequencing library preparation:** Droplets from the sorting experiment were collected in one tube and frozen at -20. Excess oil was removed with a syringe and thin needle on dry ice (to keep the sorted sample, now a small ice crystal, from entering the syringe). At ambient temperature, 10 – 30µl DNAse free water and 4-10µl PFO (1H,1H,2H,2H-Perfluoro-1-octanol, Sigma-Aldrich), was added to collect the sample, dilute the salt concentration and break the emulsion (remove most excess oil). Next, the biotinated PCR amplicons were removed from the sample by three washes on magnetic Streptavidin C1 Dynabeads (Invitrogen). The binding protocol was used according to the manufacturer's instructions, with an DNA immobilisation incubation time of 15min each run to ensure binding of long amplicons at low DNA concentrations. To collect genomic DNA and not the amplicons, we collected the supernatant of the binding assay and disposed the beads.

The amplicon-cleaned genomic DNA was desalted and cleaned by binding to magnetic SPRI beads (Beckman Coulter) (0.95x), and then the NebNext ULTRA II kit (New England Biolabs) protocol was followed for Illumina sequencing library preparation, increasing incubation times three-fold to account for low DNA concentrations. This protocol involves the following steps: DNA fragmentation end repair, adaptor ligation (adapters need to be diluted 25-fold due to very low DNA concentration), SPRI bead clean-up without size selection, PCR enrichment of fragments (equivalent to whole genome amplification, 13 PCR cycles used here), and SPRI bead clean-up without size selection. Before and after amplification, the DNA was quantified with qBit (Invitrogen) and Bioanalyzer (Agilent), initially proving the amplicon removal (no fragments at all will most likely be visible at this step due to low concentrations) and later showing the DNA fragment size distribution. The genomic DNA fragments during droplet PCR and was usually in a good size range for sequencing, however, further fragment size selection with SPRI beads might be necessary in particular when several barcoded samples have to be pooled. At last, the sample or sample-pool was sequenced with a 150 or 250 paired end kit on a MiSeq instrument (Illumina).

**Sequencing based assessment of target bacterial enrichment and genome coverage:** Reads obtained from the shotgun metagenomic sequencing of 8 metagenomic samples were quality-filtered by removing reads shorter than 70% of the maximum expected read length (100 bp), with an observed accumulated error >2 or an estimated accumulated error >2.5 with a probability of ≥0.01[28], or >1 ambiguous position. Reads were trimmed if base quality dropped below 20 in a window of 15 bases at the 3' end, or if the accumulated error exceeded 2 using the sdm read filtering software[29]. To obtain the coverage profile over the target *B. subtilis* genome, filtered reads were mapped using Bowtie2 v 2.3.4.1[30], with the parameters "--no-unal --end-to-end --score-min L,-0.6,-0.6". Resulting bam files were sorted, duplicates removed and indexed using Samtools 1.3.1[31]. Reads mapping with a mapping quality < 20, <95% nucleotide identity or <75% overall alignment length were filtered using custom Perl scripts. From these depth profiles were created using bedtools v2.21.0 which were translated with a custom C++ program "rdCov" (https://github.com/hildebra/rdCover) into average coverage in a 50 bp window. To estimate the abundance of different genera, Kraken[32] profiles were created for these samples, using default parameters and the default databases „minikraken" available from https://ccb.jhu.edu/software/kraken/. Abundance and genome coverage profiles were visualized in R 3.6.2.

**Genome sequencing based relative abundance assessment:** Sequencing results were first checked using FastQC v0.11.8[33]. Relative abundance of mixed bacterial culture samples was calculated by mapping reference genomes for *Bacteroides vulgatus* PC510 (ADKO01000001.1- ADKO01000117.1), *Bacteroides vulgatus* ATCC 8482 (UYXB01000001.1- UYXB01000005.1) and *Bacillus subtilis subsp. subtilis* str. 168 BS168_ctg (ABQK01000001.1- ABQK01000005.1) against all for samples. Mapping was done using the BWA v0.7.15 [34] commands 'bwa mem' and Samtools v1.7[31] command 'samtools view -hb' and 'samtools sort'. Generated bam files were compiled using the script 'jgi_summarize_bam_contig_depths –outputDepth depth.txt *.bam'. Contig length calculated in Depth.txt were summed to get genome size. Read abundance for each contig was averaged to calculated average read abundance. Average read abundance was divided by genome size to normalized and multiplied $1\times10^7$ by for comparison between samples.

**Genome assembly and analysis:** Sequencing results were first checked using FastQC v0.11.8[33]. Both sorted endogenous *B. vulgatus* reads enriched from a stool sample were trimmed using the commands 'interleave-fastq'[35] and 'sickle pe -c interleaved.fastq -t sanger -m combo.fastq -s singles.fastq'[36]. Samples were then individually assembled using the Spades v3.14.1 program[37] with the command 'spades.py --12 combo.fastq -s singles.fastq -o output'. Assembled contigs for both samples were mapped to both sorted stool samples using BWA and Samtools commands previously described under 'Genome sequencing based relative abundance assessment'. Generated bam files were compiled using the script 'jgi_summarize_bam_contig_depths –outputDepth depth.txt *.bam' and then binned using the metabat v2:2.15 command 'metabat -i scaffolds.fasta -a depth.txt --seed 1987 -o output –unbinned' [38]. Completeness and Contamination of resulting metagenomic assembled genomes (MAGs) was determined by CheckM v1.0.13[39] using the 'lineage_wf' flag.

# Results

**Absolute cell counts of microbiome samples give measure of cell density for encapsulation**

The density of cells in here-processed stool samples were quantified with flow cytometry, in order to ensure that the samples can be diluted to a reproducible density (here ca. 133,000 cells per microliter) suitable for single cell encapsulation into monodisperse droplets. To achieve volumetric microbial density measures, the stained and diluted samples were mixed with reference beads (polystyrene colloids) of a known concentration. The reference bead count in the flow cytometric analysis then indicated the analysed volume of sample suspension. Results are shown in Figure 2A, generated according to the method detailed under "Stool sample preparation and cell counting". Flow cytometric analysis with unspecific DNA stains has previously been shown to be an appropriate analysis tool for such complex samples[40,41]. The cell density between stool samples varied roughly within the same order of magnitude, see Figure 2A. In particular sample "#e", the only infant stool sample used, had a low cell number per volume compared to adult samples, which agreed with our expectations. The cell density in stool samples was much higher than that of typical liquid cultures.

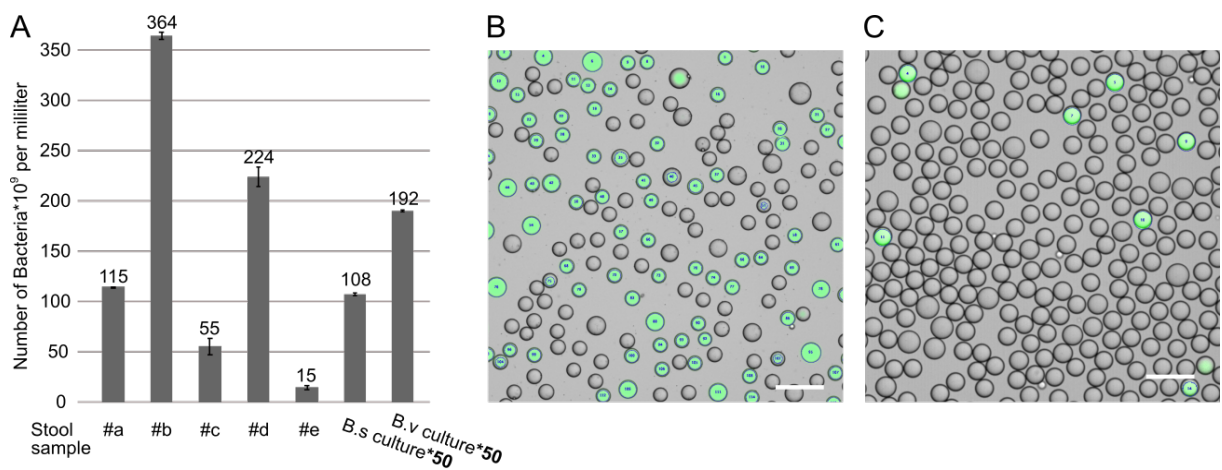

*Figure 2:* Bacterial density based on flow cytometry measurements, and microscopy images of emulsions. **(A)** Number of SytoBC-stained cells counted per volume of stool sample before dilution, and two overnight culture aliquots of bacterial cultures *Bacillus subtilis* "B.s." and *Bacteroides vulgatus* "B.v.", here shown increased by a factor of 50 to visualise it despite

of the lower cell densities compared to stool. Measurements were performed in triplicates. **(B,C)** Microscopy images of droplet emulsion containing diluted *B. subtilis* culture post droplet PCR. The images are overlays of the microscope's fluorescence and brightfield channel as well as the contours of positive-fluorescent areas, as returned from the droplet analysis tool described in the methods section. The scale bars correspond to 100µm.

**Primer-pairs and molecular probes can be made specific to target species**

In order to target any microbial strain from a sample of interest, with or without a reference genome, we introduced the use of marker gene variants of the meta mOTUs V2.0[24] tool for primer and probe design. Single copy phylogenetic marker gene (MG)-based Operational Taxanomic Units (mOTUs) quantitatively analyses the relative abundance of taxa in the sample by mapping shotgun metagenomic data to single-copy, non-16s ribosomal RNA marker genes. Marker gene sequences were obtained directly from shotgun metagenomic sequencing data, without requiring reference databases. Specific to these sequences, we designed primer pairs and, as a second layer of specificity, also TaqMan probe sequences. After designing several alternative primers and probes, they were shortlisted experimentally through testing in qPCR assays, gel-electrophoresis and sanger sequencing of PCR products (see methods "Primer and probe design and testing"). Finally, in microfluidic droplets, the TaqMan probe fluorescence lit up as expected for the corresponding bacterial culture, see Figure 2B&C, and specifically for the few spike-in culture cells in the presence of many other alternative bacterial sequences in stool, see Figure 3D-F, as expected from earlier qPCR benchmarking. The best performing sequences for our target organisms are given in Table 1.

**Stoichiometry of genetic single-cell bacteria assay in microfluidic droplets**

The number of cells per droplet follows a Poisson distribution and cannot be controlled tightly with passive encapsulation. To conduct single cell experiments in droplets, the cell suspension of known density was therefore generally diluted to a ratio of one cell in four volumes of a monodisperse droplet (here chosen at ca. 30pl) to maximize the fraction of droplets hosting single cells. This volume dilution ratio (0.25) corresponds to the Poisson parameter ($\lambda$) of the cell distribution in droplets. At this droplet occupancy ratio, the double-encapsulation of suspended cells into the same droplet is rare (2.4%)[42].

To estimate the efficiency of the PCR assay, in other words to which degree positive droplet ratios corresponded to theoretical occupancy values, we here encapsulated *Bacillus subtilis* culture stock at a lower and a higher dilution ($\lambda$ = 1 and 0.0625) compared to standard experiments. To reach desired dilution ratios, an adjusted volume of *Bacillus subtilis* cell suspension was added to the droplet encapsulation PCR mix as detailed in Methods Table 2, given the culture stock cell density as determined by flow cytometry (ca. two million cells per microliter, see Figure 2A). We then assessed the positive droplet ratio of resulting emulsions post PCR, on the basis of a custom ImageJ macro, see methods "Microscopy and evaluating positive droplet ratios". The observed positive ratios were 48% ±6% (1915 out of 3970 completely imaged droplets in total across several fields of view, see representative image section in Figure 2B) and 5% ±1% (243 out of 5105 droplets, Figure 2C) respectively. The Poisson distribution returns theoretical percentages of 63% and 6% of occupied droplets for these cell encapsulation parameters, matching the observed values well and indicating an assay efficiency of 80%.

**Single microbiota cells can be identified and sorted in droplets**

Cells of the laboratory isolate of soil bacterium *Bacillus subtilis*, which are not found naturally in the gut, were spiked into stool samples at three different ratios of 1 in 50, 1 in 100, and 1 in 250 cells and encapsulated into microfluidic droplets. Given the low mixing ratio and the droplet occupancy of about one cell per four droplets, the positive ratio of the PCR assay could not be quantified by microscopy. However, we did observe rare positive droplets as expected, see Figure 3A-F.

We then sorted a total of seven *B. subtilis* spike-in emulsions, each belonging to one of three dilution ratios, to enrich target droplets. For this purpose, droplets were reinjected into the sorting chip (Figure 1 right-top) and deflected by dielectrophoresis into the collection channel of the chip (Figure 1 right-centre) whenever their signal fell within our fluorescence thresholds in terms of fluorescence intensity and signal length (Figure 3A-C). When inspected by microscopy, the sorted positive emulsion mostly showed exclusively positive droplets (Figure 1 right-bottom) but the exact ratio was difficult to assess as many droplets merged or split downstream of the sorting junction, destabilised by the electric field of the sorting electrodes. The relative number of positive droplets inside the sorting gates decreased as expected from higher spike-in ratios (Figure 3A) to lower ratios (Figure 3C). It should be noted that droplet emulsions became less monodisperse during thermocycling, and that the droplet sizes separated out partially in the syringe given the long sorting timeframes, so that the average droplet size decreased over time and sorting boundaries needed to be adjust dynamically every few hours due to this size drift as well as temperature changes affecting the laser and detectors.

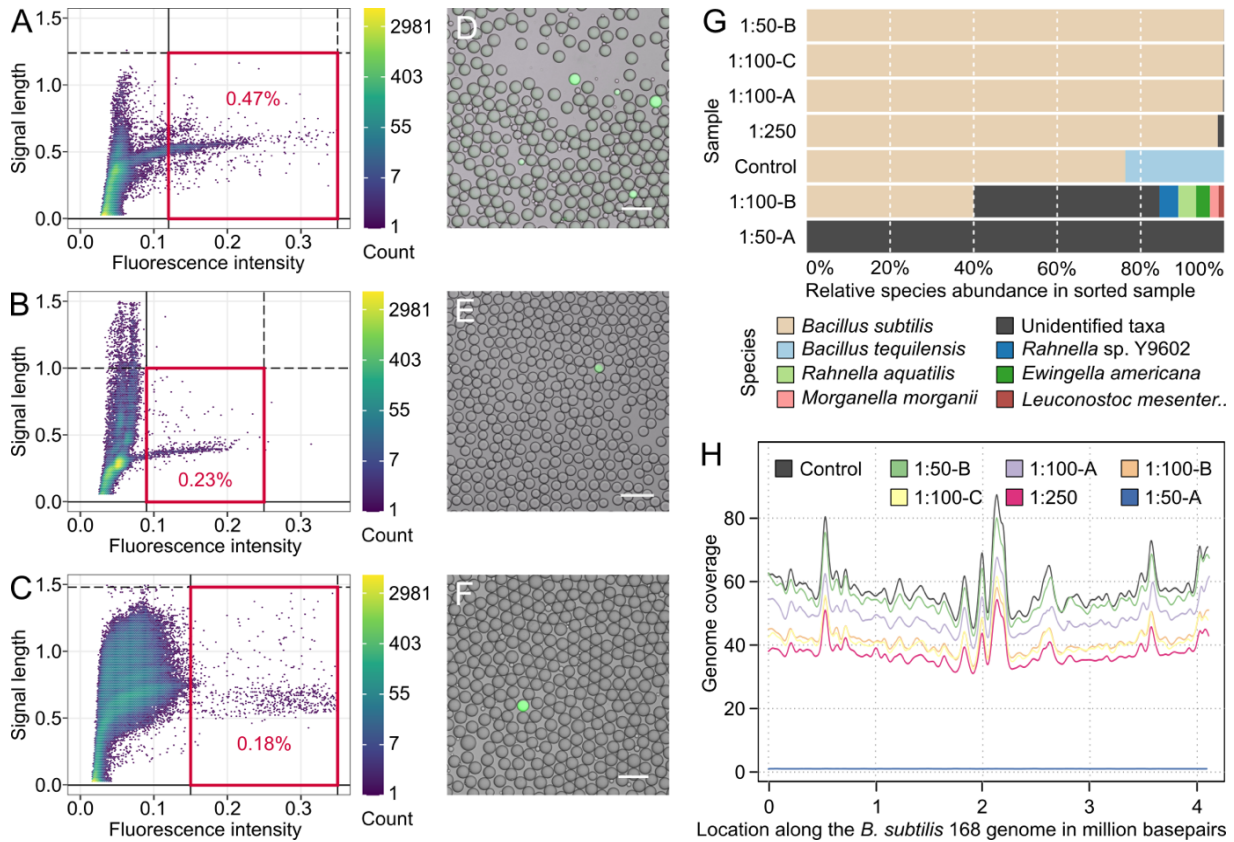

*Figure 3:* Fluorescence activated single-cell droplet sorting of *B. subtilis* cells spiked into stool samples at different ratios, of an emulsion with one in four droplets occupied by a cell. **(A,D)** About one *B. subtilis* cell in 50 microbiota cells, at 25% droplets occupied, means 1 pos. droplet in ca. 200, ratio 0.005; **(B,E)** 1:100 cells, droplet ratio 0.0025; **(C,F)** 1:250 cells, droplet ratio 0.001. **(A-B)** Density plots of droplet sorting data with the applied sorting gates highlighted in red and with percentage of events within the gate. Sections of data are shown as these experiments were run over long timeframes (ca. 10-20h of sorting per sample) and gate boundaries had to be adjusted every few hours due to temperature changes and slightly varying droplet sizes. Repeated measurements yielded similar results. **(D-F)** Microscopy overlay images of fluorescence and brightfield channel, showing positive droplets (containing *B. subtilis* culture) and negative droplets (empty or stool microbiota) post droplet PCR. The scale bars correspond to 100μm. **(G)** DNA sequencing data results of *B. subtilis* culture spiked into stool samples. The target read abundance is shown by sample. The unknown fraction (grey) could not be assigned to any bacterial genomes in our database, as further confirmed with Kraken assignments. **(H)** Mapping of target DNA sequencing reads to the *B. subtilis* reference genome, showing the genome coverage of different samples.

**Amplicons can be removed from low abundance genomic DNA**

As a byproduct of the PCR based species detection, amplicons enrich and become by far the most abundant DNA fragments inside droplets, while a droplet only contains a single copy of genomic DNA. Sorted positive droplets are thus dominated by the already known amplicon sequence. We therefore also developed a workflow for amplicon clean-up from sorted and pooled positive droplets while maintaining the low-abundant target genomes.

We used biotinylated primers for PCR, to incorporate a chemical handle into amplicons with which they could be selectively bound to magnetic Streptavidin beads, to remove them from un-labelled genomic DNA, as illustrated in Figure 1 bottom and detailed in method "Amplicon removal and sequencing library preparation". After three rounds of Streptavidin bead purification and consecutive genomic DNA re-concentration on unselectively binding SPRI beads, amplicons were reduced to an undetectable amount, see electropherograms in Supplementary Figure 1. Even after DNA amplification during Illumina sequencing library preparation, when genomic DNA become abundant, the amplicon concentration was very low or absent from recorded electropherograms. The electropherograms also revealed that the genomic DNA, molecules of several mega base pairs, already fragmented into pieces of ca. 300 to 3000 base pairs during thermocycling in droplets. There was no need for additional DNA fragmentation steps before preparing the sequencing library.

**Whole genome sequencing output provides a rough estimation of bacteria species abundance**
In a simple mixing experiment, we tested the quantitative accuracy of sequencing read counts as a prediction of input cell number ratios, as a proxy for later purity assessments of DNA libraries prepared from sorted droplets. For this purpose, we mixed bacteria cultures *Bacteroides vulgatus* PC510 strain and *Bacillus subtillis* strain 168 BS168_ctg based on their cell density as determined by flow cytometry. Cells were mixed at a ratio of (i) 10:1 and (ii) 1:1, thermocycled in droplets and sequencing libraries prepared in accordance to our overall workflow, see method "Amplicon removal and sequencing library preparation", but without droplet sorting to enrich a target. Sequencing resulted in 376,322 and 441,707 high quality paired reads for sample (i) and (ii) respectively. Reads of the two mixed of *B. vulgatus* and *B. subtilis* culture samples were mapped to their reference genomes (see methods) to calculate their relative abundance in each sample. *B. subtilis* had higher relative abundance in both samples with the ratio of *B. vulgatus* to *B. subtillis* 286:33 in sample (i) and 253:201 in (ii), revealing an average relative divergence to the original mixing ratio of 21%.

Alongside *Bacillus subtilis*, *Bacteroides vulgatus* culture was used in this particular benchmark because its wild strains also frequently occur in the gut and such endogenous strains were targeted during droplet sorting of later experiments.

**Sequencing of enriched samples shows high target abundance**
With the above error rate of sequencing-based cell-ratio quantification in mind, we sequenced the seven sorted spike-in samples discussed in section "Single microbiota cells can be identified and sorted in droplets" and binned recovered sequences to estimate the target enrichment in experiments. This typically showed a high enrichment level with >98% of recovered reads being assigned to *B. subtilis* (4/7 samples, see Figure 3G). However, some samples showed DNA reads of other origin. E.g. sample "1:50-A" had extremely few reads overall (78,000 read pairs, the average of all seven samples being 1,640,780 read pairs), indicating insufficient levels of input DNA, and reads from this sample were not assigned to any taxa (including no *B. subtilis*) by the Kraken pipeline.

The spike-in experiments also demonstrated which minimum number of sorted droplets of the target species (corresponding roughly to cells at the beginning of library preparation) is necessary to recover a high-quality genome, which ultimately limits our overall assay sensitivity in terms of target strain abundance, given the ability to sort a maximal number of droplets per experiment. Indeed, all samples except "1:50-A" resulted into a relatively even genomic coverage of *B. subtilis* (Figure 3H), indicating that an assembly of this genome is possible based on our recovered reads. It is of note, that a few regions seemed to be recovered at a higher rate in all experiments independently. These could represent duplicated genome regions and they did not overly imbalance the experiment. The good target genome coverage even by our most diluted sample (1:250 spike-in) indicates that only about 4000 sorted target cells (from ca. 1 million droplets) were needed to recover sufficient reads of a target genome.

**High-quality de-novo genomes of endogenous microbiota bacteria were assembled**
After the above extensive spike-in benchmarking of the method, we applied our precision genomics approach to the endogenous stool bacterium *Bacteroides vulgatus* to enrich and sequence its genome de-novo. The targeted experiment was performed twice using aliquots from the same stool sample, with PADS data and images shown in Figure 4B. No bacterial cultures where involved. Sequencing of the two final genomic DNA libraries, together occupying half an Illumina MiSeq run, resulted in (1) 460,513 and (2) 1,045,142 high quality paired reads respectively. Mapped read abundance for sorted stool samples showed high relative abundance of two endogenous *Bacteroides vulgatus* strains PC510 and ATCC 8482 in the total pool of sequencing reads. The read abundance for PC510 was (1) 262 and (2) 603 and for ATCC 8482 (1) 116 and (2) 332. Assembly of reads resulted in a total assembly length of 11,200,633 base pairs (bp) and 19,045 scaffolds with 526 contigs >2500 bp for sample (1). Sample (2) assembled 36,865 scaffolds with a total length of 21,451,804 bp and 708 contigs >2500 bp.

Binning of these scaffolds >2500 bp resulted in two high quality MAGs, one from each sample designated as B_vulg_1 and B_vulg_2. We performed genome quality benchmarks on our sequencing results using available reference genomes of the strains to which our reads mapped. Taxonomy, Completeness and Contamination was determined by CheckM[39]. CheckM's bacteroidales marker set calculated B_vulg_1 as 88.78% complete with 0.56% contamination and B_vulg_2 as 99.25% complete with no contamination. One medium quality MAG (70% complete) was also binned from sample (2) belonging to the family of Enterobacteriaceae. The statistics and coverage of the two assembled target genomes in comparison to the reference strain ATCC 8482 of the same species are summarised in Figure 4A&C. Larger mapping gaps (Figure 4C) can most likely be attributed to local sequence differences between the strain encountered in the donor vs. the reference strain.

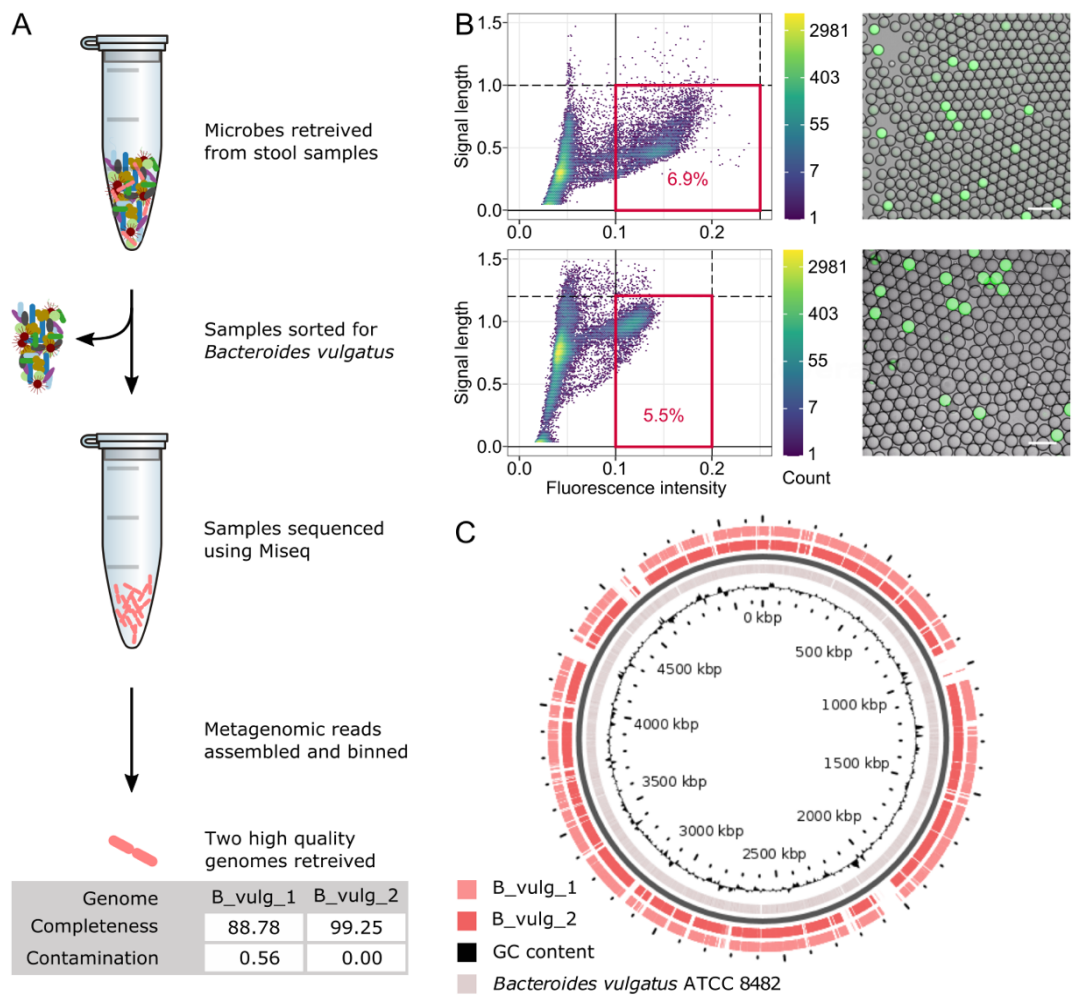

*Figure 4:* Workflow summary of the precision genomics approach to enrich target endogenous species from its microbiome community with microfluidic droplets and summary of the two high quality *Bacteroides vulgatus* genomes assembled in our experiments. **(A)** Workflow schematic and genomes statistics. **(B)** Two biological replicate experiments of fluorescence activated single-cell droplet sorting of endogenous *Bacteroides vulgatus* cells in stool samples, of an emulsion with about one in four droplets occupied by a cell. **(B Left)** Density plots of droplet sorting data with the applied sorting gates highlighted in red and with percentage of events within the gate. Sections of data are shown. **(B Right)** Microscopy overlay images of fluorescence and brightfield channel, showing positive droplets (containing endogenous *B. vulgatus*) and negative droplets

(empty or other stool microbiota) post droplet PCR. The scale bars correspond to 100μm. **(C)** The two assembled genomes were mapped to a reference genome using the blastx algorithm, illustrating (https://server.gview.ca/) the substantial genome coverage achieved here.

## Discussion

We have demonstrated how the same frozen gut microbiome (stool) samples of previous metagenomic analyses can be prepared for single cell genetic assays in microfluidic droplets. A bacteria cultivation free workflow using complex gut microbiome samples was demonstrated and benchmarked with in-droplet digital PCR, droplet sorting, genomic library preparation and sequencing data analysis. Evidence has been provided that single-cells can be targeted, endogenous target species can be enriched through droplet sorting, amplicon sequences from the genomic DNA can be removed, and whole genomes can be sequenced to obtain high-quality and low-contamination drafts of the strains of the target bacterial species. The method was applied to the endogenous gut species *Bacteroides vulgatus*, and two high-quality, low-contamination genomes were recovered (without using or requiring reference genomes) from the same stool sample. We believe this method to be an important resource for the scientific community in their mission to advance from cell cultures to apply precision genomics methods directly to complex samples of medical or biotechnological relevance. In the following paragraphs, performance and limitations in comparison to other methods are discussed.

Typically, molecular DNA probes are designed to target a variable region of the 16S RNA marker gene e.g.[43,44], based on sequences in reference databases or amplicon sequence variants (ASVs) in samples[45]. However, the variable regions of the 16S RNA gene are short and therefore provide little space for highly specific probe design. Accordingly, 16S variable region sequences as routinely processed with Illumina sequencing do not allow to distinguish between species or strains of interest. In contrast, the single-copy marker gene approach, meta mOTUs v2[24], enables a fast species level assessment of the microbial diversity and relative abundances of a microbiome sample even without reference databases. It uses ten marker genes which are longer than the 16S-RNA gene. Besides advantages in resolution, relative abundances based on these single-copy genes are also more accurate than those based on the 16S RNA gene, because the latter can be present in multiple copies per cell. In this work, we successfully demonstrate the reference-free use of meta mOTUs to design specific primers and molecular probes directly on shotgun metagenomic data from a sample, specifically on the comparably large sequence space of the ten marker genes within the metagenomic data. The meta mOTUs accuracy of relative abundances furthermore contributes valuable information when assessing if target species have a sufficient abundance in the sample to yield a high-quality result given the experimental power available. This method can be directly applied to new targets also in other habitats.

The best known culture-free method to obtain individual genomes from a microbiome sample is the generation of single amplified genomes (SAGs)[18,19] from single cells. A few hundred genomes can be recovered in parallel with low contamination when using high-throughput droplet microfluidic methods for SAGs, and the most abundant species may be represented with several SAGs that can be pooled to increase genome quality[18]. Since this method is not targeted, however, it is not certain that any specific genome is among the results, and the genomes have a low average quality[46]. The here-presented targeted approach is, in contrast, more likely to recover a high-quality genome of the target species, thereby complementing existing SAG methods. In this precision genomics resource we show with sequencing data of spike-in bacterial culture, that bacteria can be enriched sufficiently even at spike-in ratios as low as 1:250, with as little as 4000 enriched target cells, to achieve a ca. 40x coverage of the target genome, which is suitable for a reliable genome assembly.

Absolute quantification of bacterial load and the abundance of target species is an important complementary information for metagenomic data[40,41]. The droplet digital PCR (ddPCR) approach has the advantage that it can be used for the absolute enumeration of targeted bacterial cells within a volume of complex environmental sample, in addition to recovering the target genomes later on as shown here. ddPCR has previously been used for the quantitative detection of various bacterial pathogens[47], DNA copy number[48], and viral sequence abundances[49], even outperforming qPCR assays[50,51]. For absolute cell quantification by ddPCR, an estimate of the assay efficiency is needed to conclude the cell number from an observed positive droplet ratio. We conclude from our data that the assay provides a visual readout of approximately 80% of the theoretical number of droplets occupied with target cells (see results "Stoichiometry of genetic single-cell bacteria assay in microfluidic droplets"), this efficiency was stable between different samples and bacteria tested. There are a number of experimental factors which limit the assay efficiency to below 100% and cause some level of variation between experiments: (i) temperature-only based cell lysis efficiency during thermocycling and (ii) PCR efficiency <100%[21], (iii) manual fluorescence intensity thresholding setting against background, and (iv) error in bacterial density measurement and dilution of input culture.

It has been shown that microfluidic droplets are particularly well suited to cultivate microorganisms[52,53], including slow growing species. Nevertheless, the majority of bacteria and even more so archaea[54] have never been successfully cultivated in the laboratory[55,56]. We therefore developed a cultivation free resource which can be directly applied to other habitats. The method does not even rely on purifying cells or keeping them alive, which allowed us e.g. to analyse samples stored for months in -80 freezers and to work with anaerobic gut bacteria under standard aerobic laboratory conditions. Available cultures and reference genomes do help, however, to benchmark primer and probe designs for their specificity, to ensure the lysis efficiency of cells, and to compare results.

While we developed this method, significant progress has been made in purely computational tools to assemble genomes from metagenomic datasets[57,58]. Such metagenomically assembled genomes (MAGs) could serve as a helpful complementary resource to SAGs and genomes recovered with this resource. On the one hand, the vast number of available MAGs can help to identify species of interest as well as aid to assess probe sequence specificity. Sorted genomes, on the other hand, may be used to improve the quality of special MAGs, such as contamination and chimeric sequences. Sorted genomes come with less foreign DNA which decreases the amount of contamination in the generated bins, while also simplifying and speeding up binning and assembly. Furthermore, the cell sorting physically co-enriches mobile genetic elements that can often not be clearly assigned to genomes.

Last but not least, as the field aims to move from functional predictions to ecotypes of species, we require single cell approaches[59] to combine genetic analysis with functional screens[60]. Droplet microfluidic methods allow the use of florescent probes indicating functional properties of cells as trigger for droplet sorting[61] to enrich not only for the presence of genetic target sequences as shown here, but also industrially-relevant enzymatic activities[62,63], cell-surface proteins via antibodies[11], functionalised quantum dots[64], or other probes targeting cell surfaces and bioactive content. These approaches can be combined with the presented resource to achieve powerful multi-omic readouts. But even without combination, this method allows a direct target enrichment of functional sequences, e.g. all taxa with a genetic region indicating esterase activity, particular secondary metabolites or other desirable functions, alleviating resolution problems associated with sequencing depth of untargeted libraries. The droplet sorting approach is particularly attractive to target mobile genetic elements such as viral sequences[20,49] or transposons for antibiotic resistance[65], which can often not be clearly associated in metagenomic data[66].

## Data availability
OSF https://osf.io/6tmzk/ (CheckM, Binning, Chips, PADS data plotting via GitHub)
GitLab https://gitlab.com/marco.r.cosenza/simple_droplet_tools (Droplet-analysis FIJI Macro)

## Acknowledgments
The authors thank the EMBL Genomics and Flow Cytometry core facility teams for training and advice. We are grateful to the support by Nassos Typas and his group, in particular Sarela Garcia Santamarina and Matylda Zietek, for the provision of cultures, space and advice. We also thank Alessio Milanese for discussions on the use of the meta mOTU v2 tool.

All authors are grateful for funding and support by the EMBL. AP, TW and MRC have been funded by EIPOD, the European Commission Marie Curie Program with EMBL (TW: EI3POD, grant no. 664726). FH's salary is funded by the BBSRC Institute Strategic Programme Gut Microbes and Health BB/r012490/1, its constituent project BBS/ e/F/000Pr10355. This project has also received funding from the European Research Council (ERC) under the European Union's Horizon 2020 research and innovation programme (grant agreement no. 948219). EK was funded by the European Research Council (ERC-AdG-669830 MicrobioS).

## Author Contributions

CM, PB, AP: Conceptualization; CM, PB, AP, VB, TW, EK: Methodology, Validation; AP: Investigation; CM, PB: Supervision; CM, PB, AP, TW: Funding acquisition; MRC, KWS, FH: Software; KWS, FH, AP, TW: Data Analysis; TW: Validation, Review, Writing - Original Draft, Visualisation; All: Reviewing and Editing.

## Declaration of Interests

The authors declare no competing interests.

# Supplementary Figures

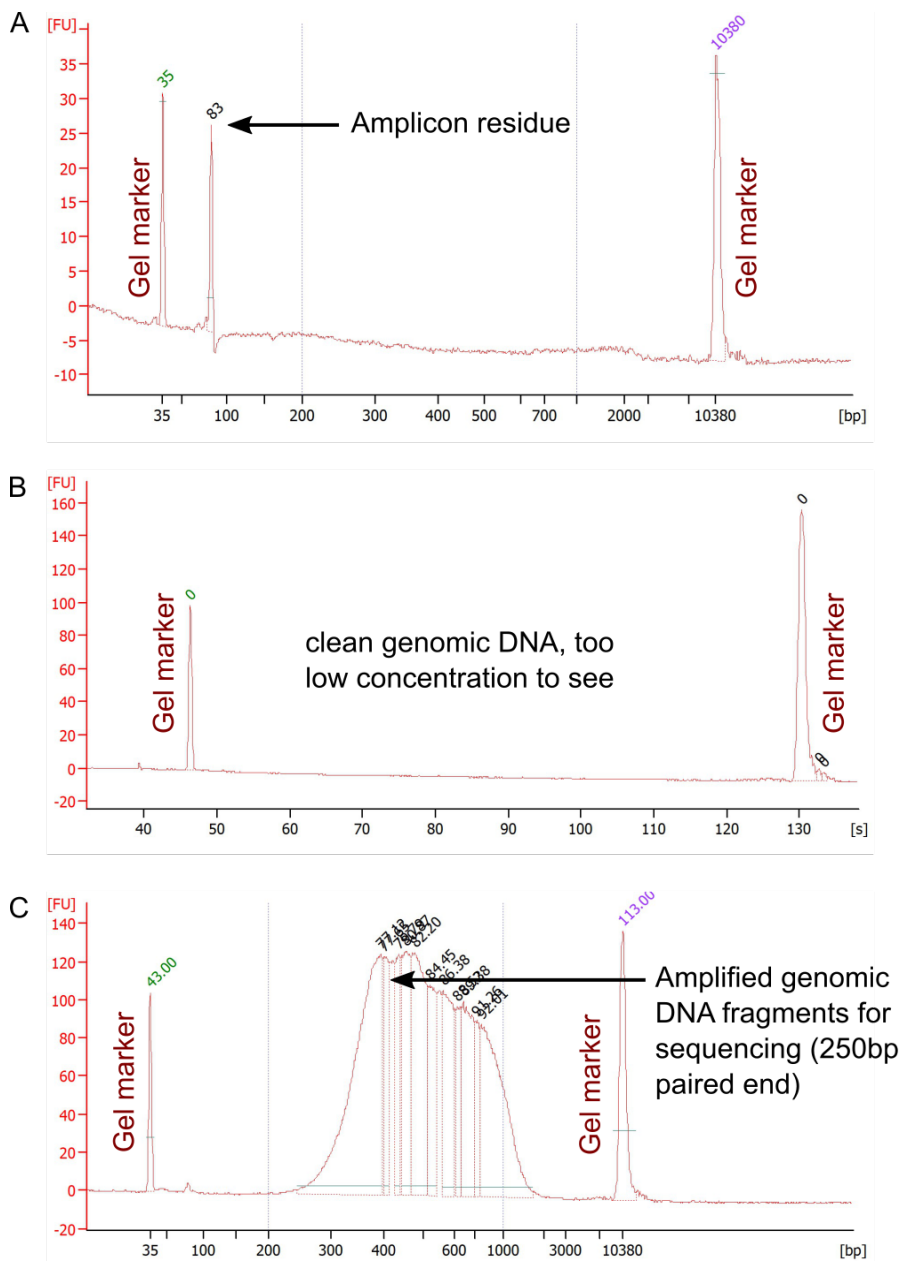

***Supplementary Figure 1:*** Electropherogram traces recorded with an Agilent Bioanalyzer. The x-axis represents the DNA fragment length in base pairs [bp] and the y-axis the amount of DNA detected of such fragments in fluorescence units [FU]. **(A)** DNA from sorted droplets after insufficient Streptavidin beads clean-up; **(B)** Clean genomic DNA that is visibly free of amplicons, but the genome fragment concentration is too low to be detected before amplification; **(C)** Cleaned, amplified (and size selected where needed) genomic sample library, ready for pooling and sequencing. Even without size selection, the genomic DNA recovered from droplets had fragment lengths mostly between 300bp and 3kbp, which indicates fragmentation of the large genome molecules (several Mbp) during thermocycling.